\documentstyle{article}
\renewcommand{\thefootnote}{*}
\begin{document}
\newcommand{\D}{{\cal D}}
\newcommand{\C}{{\bf C} \!\!\!\! I}
\newcommand{\HH}{H \!\!\! I}
\newcommand{\UU}{\underline{\sqcup}}
\newcommand{\UUU}{\sqcup}
\newcommand{\Z}{Z \!\!\! Z}
\newcommand{\la}{\langle}
\newcommand{\ra}{\rangle}
\newcommand{\li}
{\begin{picture}(10,10)(0,0)
\put(0,-3){\line(0,1){10}}
\end{picture}}
\newcommand{\sq}{\begin{picture}(10,10)(0,0)
\put(0,-3){\line(0,1){10}}
\put(-2,2){${}_{\Box}$}
\end{picture}}
%{| \!\! {}_{\Box}}
\newcommand{\bl}
{\begin{picture}(10,10)(0,0)
\put(0,-3){\line(0,1){10}}
\put(-2,0){$\bullet$}
\end{picture}}
%{| \!\! \bullet}
\newcommand{\dbl}
{\begin{picture}(10,15)(0,0)
\put(0,-3){\line(0,1){15}}
\put(-2,0){$\bullet$}
\put(-2,4){$\bullet$}
\end{picture}}
\newcommand{\dsq}
{\begin{picture}(10,15)(0,0)
\put(0,-3){\line(0,1){15}}
\put(-2,1){${}_\Box$}
\put(-2,7){${}_\Box$}
\end{picture}}
\newcommand{\cus}{\bigcup_{\!\!\!\! \Box}}
\newcommand{\cub}{\bigcup_{\!\!\!\! \bullet}}
\newcommand{\sms}{
\smile \!\! \circ}
\newcommand{\smb}{\smile \!\!
\bullet}
\newcommand{\rec}{\begin{picture}(0,0)(0,0)
\put(-15,60){\dashbox{1.0}(140,40){}}
\end{picture}}
\newcommand{\Cup}{
\begin{picture}(20,15)(-7,-7)
\put(0,0){\oval(10,15)[b]}
%\put(-3,-8){${}_{\Box}$}
\end{picture}}
\newcommand{\Cap}{
\begin{picture}(20,15)(-7,-7)
\put(0,-10){\oval(10,15)[t]}
%\put(-3,-8){${}_{\Box}$}
\end{picture}}
\newcommand{\Cus}{
\begin{picture}(20,20)(-7,-7)
\put(0,0){\oval(10,15)[b]}
\put(-3,-8){${}_{\Box}$}
\end{picture}}
\newcommand{\Cub}{
\begin{picture}(20,20)(-7,-7)
\put(0,0){\oval(10,15)[b]}
\put(-3,-10){$\bullet$}
\end{picture}}
\newcommand{\scus}{
\begin{picture}(20,20)(-7,-7)
\put(0,0){\oval(10,10)[b]}
\put(0,0){\oval(30,20)[b]}
\put(0,-10){${}_{\Box}$}
\end{picture}}
\newcommand{\scup}{
\begin{picture}(40,15)(-15,-7)
\put(0,0){\oval(10,10)[b]}
\put(0,0){\oval(30,20)[b]}
\end{picture}}
\newcommand{\scub}{
\begin{picture}(20,20)(-7,-7)
\put(0,0){\oval(10,10)[b]}
\put(0,0){\oval(30,20)[b]}
\put(0,-12){$\bullet$}
\end{picture}}
\newtheorem{th}{Theorem}
\newtheorem{cor}{Corollary}[th]
\newtheorem{de}{Definition}
\newtheorem{pr}{Proposition}
\newtheorem{co}{Corollary}[pr]
\newtheorem{rem}{Remark}

USC-93/009  February 12, 1993
\bigskip
\begin{center}
{\bf
THE
BLOB ALGEBRA AND THE PERIODIC TEMPERLEY-LIEB ALGEBRA
}
\\
Paul  Martin
\footnote{Math Dept., City University,
Northampton Square, London EC1V 0HB, UK.}
\renewcommand{\thefootnote}{\dag}
and
Hubert Saleur \footnote{Math Dept. and Physics Dept., University of Southern
California, Los Angeles, CA 90089, USA.}
\\
\end{center}

%\vspace{.2in}

%{\bf Abstract} \hspace{.3in}
\begin{abstract}
We determine the structure of two variations on the Temperley-Lieb
algebra, both used for dealing with special kinds of boundary
conditions in statistical mechanics models.
 The first is a new algebra, the `blob' algebra (the reason for the
name will become obvious shortly!). We determine both the generic and all the
exceptional structures for this two parameter algebra.
The second is the periodic Temperley-Lieb algebra. The generic structure
and part of the exceptional structure of this algebra have  already been
studied. Here we complete the analysis, using results from the
study of the blob
algebra.
\end{abstract}

\section{Introduction}
There
has been much recent interest in two dimensional
Potts models and related models in the
case of toroidal boundary conditions
\cite{McCoy,B}.
In this paper we determine the structure of two variations on the
ordinary Temperley-Lieb
algebras, which appear in the
transfer matrices of these models.

In the next section we introduce    %The first is
a new
two parameter algebra, the `blob' algebra $b_n(q,q')$ (it turns out to be a
particular quotient of the affine Hecke algebra).
Using diagrams, some ideas from category theory,
and experience gained from analysing the ordinary T-L algebras,
we determine both the generic and all the
exceptional structures
(depending on $q,q'$ and also on their relationship)
for this algebra.
In the subsequent section we analyse
%The second is
the periodic Temperley-Lieb algebra. The generic structure
and part of the exceptional structure of this algebra has already been
studied \cite{PS90,Lev,Lev1,MS}. Here we complete the analysis,
relying heavily on results from the
study of the blob
algebra.

A striking result
already emphasized in \cite{PS90} is the analogy of
certain special representations
of the periodic T-L algebra
with the
sum of representations of left and right Virasoro algebra $\sum Vir_{ra}
\overline{Vir}_{rb}$ (where the labels refer to the highest weights
$h_{ra}$, $\overline{h}_{rb}$ of the Virasoro represen\-tations) \cite{BPZ}.
We will develop the technology to examine how
 this analogy goes further, that is,  representations
that are not in the special part of the Bratteli diagram
already studied can also be
put in correspondence with representations of the Virasoro algebra, but now
``outside'' the minimal Kac table.

This work has some overlap with the papers of Levy \cite{Lev,Lev1}.
However we believe that the natural graphical representations we use provide a
clearer and more complete point of view on the representation theory.
In conclusion we also mention some interesting possibilties for further
generalizations

\section{The blob algebra}
\subsection{Definition and general results}
\noindent
%{\bf Introduction}

Recall the well known diagrammatic realisation of the Temperley-Lieb
algebra $T_n(q)$ \cite{TL,Pen} in which the generators are drawn
as $n$ non-overlapping strings on a rectangular frame

%\vspace{1in}
%\begin{figure}
\begin{picture}(100,70)(-100,40)
\put(0,80){$\huge 1 = $}
\put(60,0){
\rec
\thicklines
\multiput(0,100)(10,0){12}{\line(0,-1){40}}
\thinlines
}
\end{picture}

\begin{picture}(100,70)(-163,30)
\put(-70,70){{\Huge U}${}_{i} \;$ {\Huge =}}
%
%\put(-15,100){\line(1,0){140}}
%\put(-15,100){\line(0,-1){40}}
%\put(-15,60){\line(1,0){140}}
%\put(125,60){\line(0,1){40}}
%
\rec
\thicklines
\multiput(0,100)(10,0){5}{\line(0,-1){40}}
\put(55,100){\oval(10,20)[b]}
\put(55,60){\oval(10,20)[t]}
\put(-2,103){{\tiny 1}}
\put(8,103){{\tiny 2}}
\put(18,103){..}
\put(47,103){{\tiny i}}
\put(54,103){{\tiny i+1}}
\put(67,103){ }
\put(87,103){..}
\put(106,103){{\tiny n}}
\multiput(70,100)(10,0){5}{\line(0,-1){40}}
\thinlines
\end{picture}
%\caption{Boundary
%diagram for $U_i$ in $D_{A_n}(Q)$, c.f. $A_{j.}$ ($j=(i+1)/2$)
%in the connectivity picture. \label{Uipic}}
%\end{figure}
%

\noindent
and composition is by identification of the bottom of one diagram with the
top of the other
(the exterior rectangles are
construction lines only, and can be
ignored in composition). The relations are

\begin{picture}(100,100)(-50,10)
\put(0,0){
%\put(-15,100){\line(1,0){140}}
%\put(-15,100){\line(0,-1){40}}
%\put(-15,60){\line(1,0){140}}
%\put(125,60){\line(0,1){40}}
\rec
\thicklines
\multiput(0,100)(10,0){5}{\line(0,-1){40}}
\put(55,100){\oval(10,20)[b]}
\put(55,60){\oval(10,20)[t]}
\multiput(70,100)(10,0){5}{\line(0,-1){40}}
\thinlines
}
%\end{picture}
%
%\begin{picture}(10,10)(-100,-10)
\put(0,-40){
%
%\put(-15,100){\line(1,0){140}}
%\put(-15,100){\line(0,-1){40}}
%\put(-15,60){\line(1,0){140}}
%\put(125,60){\line(0,1){40}}
\rec
\thicklines
\multiput(0,100)(10,0){5}{\line(0,-1){40}}
\put(55,100){\oval(10,20)[b]}
\put(55,60){\oval(10,20)[t]}
\multiput(70,100)(10,0){5}{\line(0,-1){40}}
\thinlines
}
%\end{picture}
%
%\begin{picture}(1,1)(200,30)
\put(130,60){$\huge = \; x$}
\put(200,-10){
%
%\put(-15,100){\line(1,0){140}}
%\put(-15,100){\line(0,-1){40}}
%\put(-15,60){\line(1,0){140}}
%\put(125,60){\line(0,1){40}}
\rec
\thicklines
\multiput(0,100)(10,0){5}{\line(0,-1){40}}
\put(55,100){\oval(10,20)[b]}
\put(55,60){\oval(10,20)[t]}
\multiput(70,100)(10,0){5}{\line(0,-1){40}}
\thinlines
}
\end{picture}

\noindent
(where $x=q+q^{-1}$ and $q$ refers to the usual $U_qsl(2)$ quantum algebra)
and equivalence under end point preserving isotopy, e.g.

%\vspace{1in}
\begin{picture}(100,140)(-50,-25)
\put(0,0){
%
%\put(-15,100){\line(1,0){140}}
%\put(-15,100){\line(0,-1){40}}
%\put(-15,60){\line(1,0){140}}
%\put(125,60){\line(0,1){40}}
%
\rec
\thicklines
\multiput(0,100)(10,0){5}{\line(0,-1){40}}
\put(55,100){\oval(10,20)[b]}
\put(55,60){\oval(10,20)[t]}
\multiput(70,100)(10,0){5}{\line(0,-1){40}}
\thinlines
}
%\end{picture}
%
\put(0,-40){
%
%\put(-15,100){\line(1,0){140}}
%\put(-15,100){\line(0,-1){40}}
%\put(-15,60){\line(1,0){140}}
%\put(125,60){\line(0,1){40}}
\rec
\thicklines
\multiput(0,100)(10,0){6}{\line(0,-1){40}}
\put(65,100){\oval(10,20)[b]}
\put(65,60){\oval(10,20)[t]}
\multiput(80,100)(10,0){4}{\line(0,-1){40}}
\thinlines
}
\put(0,-80){
%
%\put(-15,100){\line(1,0){140}}
%\put(-15,100){\line(0,-1){40}}
%\put(-15,60){\line(1,0){140}}
%\put(125,60){\line(0,1){40}}
\rec
\thicklines
\multiput(0,100)(10,0){5}{\line(0,-1){40}}
\put(55,100){\oval(10,20)[b]}
\put(55,60){\oval(10,20)[t]}
\multiput(70,100)(10,0){5}{\line(0,-1){40}}
\thinlines
}
\put(140,40){$\huge = $}
\put(200,-30){
%
%\put(-15,100){\line(1,0){140}}
%\put(-15,100){\line(0,-1){40}}
%\put(-15,60){\line(1,0){140}}
%\put(125,60){\line(0,1){40}}
\rec
\thicklines
\multiput(0,100)(10,0){5}{\line(0,-1){40}}
\put(55,100){\oval(10,20)[b]}
\put(55,60){\oval(10,20)[t]}
\multiput(70,100)(10,0){5}{\line(0,-1){40}}
\thinlines
}
\end{picture}

\noindent
It is convenient to refer to lines which travel from top to bottom
as {\em propagating} lines, and those that double back to the same edge as
{\em loop}
lines (so, for example, the pictures above each have ten propagating lines).

The following generalisation has several applications, which we will
discuss later.

\newpage

For $q'$ an invertible complex parameter we define the BLOB algebra
$b_n = b_n(q,q')$ as the generalisation obtained by including an additional
idempotent `blob' generator

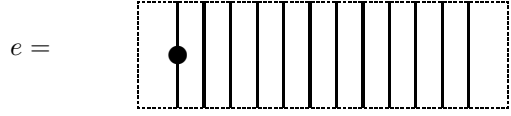
\begin{picture}(100,100)(-100,30)
\put(0,80){$\huge e = $}
\put(60,0){
%
%\put(-15,100){\line(1,0){140}}
%\put(-15,100){\line(0,-1){40}}
%\put(-15,60){\line(1,0){140}}
%\put(125,60){\line(0,1){40}}
\rec
\thicklines
\multiput(0,100)(10,0){12}{\line(0,-1){40}}
\put(0,80){\circle*{7}}
\thinlines
}
\end{picture}

\noindent
and additional relations given by $\;$
\bl $= \;$  \dbl (idempotency) and

%\newpage

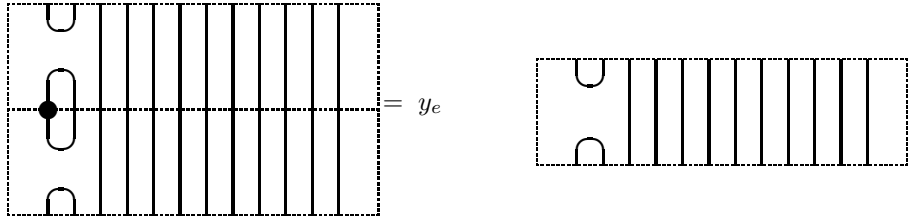
\begin{picture}(100,100)(-50,10)
%
%\put(-15,100){\line(1,0){140}}
%\put(-15,100){\line(0,-1){40}}
%%\put(-15,60){\line(1,0){140}}
%\put(125,60){\line(0,1){40}}
%
\put(0,0){
\rec
\thicklines
%\multiput(0,100)(10,0){5}{\line(0,-1){40}}
\put(5,100){\oval(10,20)[b]}
\put(5,60){\oval(10,30)[t]}
\multiput(20,100)(10,0){10}{\line(0,-1){40}}
\thinlines
%\end{picture}
%
\put(0,60){\circle*{7}}
}
\put(0,-40){
%
%%\put(-15,100){\line(1,0){140}}
%\put(-15,100){\line(0,-1){40}}
%\put(-15,60){\line(1,0){140}}
%\put(125,60){\line(0,1){40}}
\rec
\thicklines
%\multiput(0,100)(10,0){5}{\line(0,-1){40}}
\put(5,100){\oval(10,30)[b]}
\put(5,60){\oval(10,20)[t]}
\multiput(20,100)(10,0){10}{\line(0,-1){40}}
\thinlines
}
%\end{picture}
%
%\begin{picture}(1,1)(200,30)
\put(130,60){$\huge = \; y_e$}
\put(200,-21){
%
%\put(-15,100){\line(1,0){140}}
%\put(-15,100){\line(0,-1){40}}
%\put(-15,60){\line(1,0){140}}
%
\rec
%\put(125,60){\line(0,1){40}}
%
\thicklines
%\multiput(0,100)(10,0){5}{\line(0,-1){40}}
\put(5,100){\oval(10,20)[b]}
\put(5,60){\oval(10,20)[t]}
\multiput(20,100)(10,0){10}{\line(0,-1){40}}
\thinlines
}
\end{picture}

\noindent
where $y_e=q'+q'^{-1}$ (the parameter $q'$ is just introduced here by analogy,
we do not know any special meaning for it as in the case of $q$).

The set of (equivalence classes of)
diagrams produced is denoted $B^n_{\bullet}$.
In addition to the original
Temperley-Lieb diagrams this now includes
diagrams with `decoration' of various lines by (single)
blobs, with the constraints that
no line to the right of the leftmost propagating line
%(i.e. line travelling from top to bottom)
 may be so decorated;
and to the left of it
only the outermost line in any nested formation of loop lines at top or bottom
may be so decorated. Equivalently, only lines exposed to
(bounding the same region of the interior of the rectangle
as) the west face of
the rectangle may be so decorated.

For symmetry
it is convenient to introduce a generator
\[
f=1-e
\]
 and scalar $y_f=x-y_e$ and represent $f$ as a
propagating
line carrying a square box, with $\;$
\sq = $\;\;$ \dsq (c.f. the circular blob for $e$). By linearity
the same space is then spanned by
the set $B^n_{*}$ of diagrams in which every line bounding
%
%PAUL, IS THIS NOTATION SAFE? WHY DOES B HAVE ONLY A BOX WHILE IT STANDS FOR
%DIAGRAMS THAT HAVE BOX OR BULLET?    ----    I THINK IT SAFE BECAUSE THE
%%OBVIOUS
% PURE BOX PICTURES ARE RELATED TO PURE BULLET BY e <-> f SYMMETRY
% (THEREFORE WE NEVER NEED THEM, OR NOTATION FOR THEM). HOWEVER, I UNDERSTAND
% THE OBJECTION, SO I HOPE YOU ARE HAPPY WITH THIS NEW ALTERNATIVE
%
the same region of the interior
as the west face of the rectangle carries either a blob or
a square.

We will give an explicit construction for the elements of $B^n_{*}$
in proposition~\ref{P}.

\begin{de}[Index of a
diagram] For $D$ such a diagram the index $|D|$ is the number of propagating
lines, and $|D|'$ is the number of propagating blobbed lines (so
$|D|'$ is either 0 or 1).
\end{de}
\begin{de}
For $h$ an integer we define $b^h_n$ as the vector space with basis
\[
B^{n,h}_{\bullet} = \{ D \in B^n_{\bullet} \; : \; \; |D| \le h \}
{}.
\]
\end{de}
\begin{pr}
\label{pr1}
For $D_1 D_2$ the algebra composition of two diagrams from $B^n_{\bullet}$
\begin{equation}
\label{e1}
|D_1 D_2 | \le min(|D_1|,|D_2|)
\end{equation}
and if $|D_1|=|D_2|=|D_1D_2|$ then
\begin{equation}
\label{e2}
|D_1 D_2 |' = max(|D_1|',|D_2|')
{}.
\end{equation}
\end{pr}
\begin{co}
As an algebra bimodule the algebra is filtered through a sequence
of invariant subspaces
\[
\emptyset \subset b^1_n \subset b^3_n \subset ... \subset b^n_n = b_n
\hspace{1in} \mbox{($n$ odd)}
\]
or
\[
\emptyset \subset b^0_n \subset b^2_n \subset b^4_n
        \subset ... \subset b^n_n = b_n
\hspace{1in} \mbox{($n$ even)}
{}.
\]
The
%se subspaces
subquotients $b^h_n/b^{h-2}_n$
in turn are each
a direct sum of a part with a propagating $e$
and a part with a propagating $f$
(apart from $b^0_n$ which has no invariant
subspaces).
%filtered in a similar way by $|D|'$
%(i.e. they each have one proper invariant subspace as bimodules,
%call it $b_k'$,
A basis for $b^{he}_n$, the
propagating $e$ part, is
$\{ D \in B^n_{\bullet}  \; : \; \; |D|=h \; \mbox{ and } \; |D|' >0 \}$).
\end{co}

%\vspace{.1in}
%
%\noindent
%IS THIS A SUBSECTION OR SOMETHING? -----  YES!
%
%{\bf
\subsection{The structure of $b_n(q,q')$}

%DO YOU REALLY MEAN REF TO WESTBURY?
%    ----  YES, I WOULD LIKE TO ENCOURAGE HIM A BIT! (I HOPE IT'S OK WITH YOU?)
%  THE ORANGE BOOK OF MARTIN ALREADY HAS ENOUGH REFERENCES.

As in the TL case \cite{Westbury} there is a diadic notation for diagrams.
Cutting each propagating line at its midpoint (for definiteness
$| \!\! \bullet$
is written $\;$ \dbl
and cut between the blobs) the diagrams may be separated uniquely
into a top part and
a bottom part, for example:

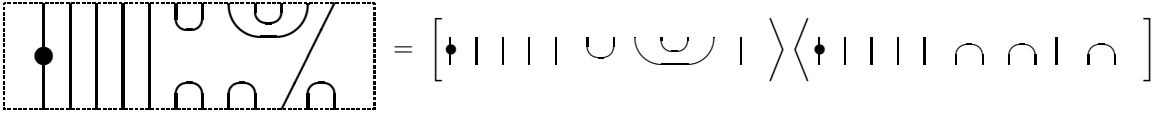
\begin{picture}(100,100)(23,30)
\put(20,0){
%
%\put(-15,100){\line(1,0){140}}
%\put(-15,100){\line(0,-1){40}}
%\put(-15,60){\line(1,0){140}}
%\put(125,60){\line(0,1){40}}
\rec
\thicklines
\multiput(0,100)(10,0){5}{\line(0,-1){40}}
\put(0,80){\circle*{7}}
\put(85,100){\oval(30,25)[b]}
\put(85,100){\oval(10,15)[b]}
\put(55,100){\oval(10,20)[b]}
\put(55,60){\oval(10,20)[t]}
\put(75,60){\oval(10,20)[t]}
\put(105,60){\oval(10,20)[t]}
\put(90,60){\line(1,2){20}}
\thinlines
}
\put(150,80){$
{
%\Huge
 \; \; = \; \left[ \;
\bl \li \li \li \li
%\; | \!\!\! \bullet \!\! ||||
%\bigcup^{\!\!\!\!\!\cup}
\Cup \scup \li \right\rangle \left\langle \;
\bl \li \li \li \li
%| \!\!\!\bullet \!\! |||| n
\Cap \Cap \li
\Cap \; \right]
}
$}
\end{picture}

Since no braiding is allowed, the recombination of such diagrams is unique
within each subquotient $b^h_n/b^{h-2}_n$.
We denote the (mutually top-bottom inverted but otherwise
isomorphic) sets of upper and lower half diagrams $R^{n,h}$,
and note that
the set $R^{n,h}$ ($h>0$)
may be split into a propagating $e$ part (called $e$-type diagrams)
$R^{n,h}_e$
and a propagating $f$ part ($f$-type) $R^{n,h}_f$.
We have
\[
B^{n,h}_{*}
\rightarrow (
R^{n,h}_e \times R^{n,h}_e)
\oplus (
R^{n,h}_f \times R^{n,h}_f)
\rightarrow
B^{n,h}_{*}
{}.
\]
We denote
 the upper and lower half diagrams extracted from a diagram $D$ by
$|D>$ and $<D|$:
\begin{equation}
D \mapsto |D> \; \; <D| \mapsto [d_1> \; <d_2] = D
\end{equation}
It is useful to make a distinction between notations $|D>$ and $[d_1>$.
The latter,
as indicated in our picture above, denotes a specific half
diagram $d_1 \in R^{n,h}$ realized as an {\em upper} half diagram.
%
%
%PAUL, CAN Y0U BE MORE PRECISE ABOUT THIS DISTINCTION? WHAT DOES $\times$
%EXACTLY MEAN HERE?
%   ------OK?
%
We will continue to use the notion of propagating lines for {\em cut}
lines, and hence of index $h$ for half diagrams with $h$ cut lines.

%First we need some notation.
For a given half diagram $d \in R^{n,h}$ we define $R_d$ as the set of all
diagrams of lower
index (actually index $\leq h-2$)
 which can be obtained from $d$ by connecting some or all
of the $h$
propagating lines in pairs to form loop lines.
%(since a propagating line cannot be trapped
%by a loop line
%\[
%|R_d|=(k-1)(k-3)(k-5)...1
%\]
%- c.f. the Pascal diagram).
In other words
\[
R_d = \{ <D| \; : \; |D|<h \mbox{ and } D \in b_n[d><d] \}
{}.
\]
Similarly define the subset
\[
R'_d =\{ <D| \; : \; |D|=h-2 \mbox{ and } D \in b_n[d><d] \}
{}.
\]

Consider the algebra product
\[
D_1 D_2 =X(D_1,D_2) \; D_3,
\]
 where $X$ is
a scalar function of $y_e,y_f$. Since the loop lines at the top of $D_1$
and bottom of $D_2$ are not affected by the product,
and $X(D_1,D_2)$ just depends on the number and nature of
closed loops produced, we have the diadic
version
%
% If we relax the $b^{k-2}$
%quotient in equation~\ref{inner} then it must be generalised to
\begin{equation}
\label{inner2}
[a><b] \; [c><d] = <b|c> \; [a'><d']
\end{equation}
where $a',d'$ are either $a,d$ or are in $R_a,R_d$ respectively.
%
%The maximal proper
An (proper) invariant subspace of the {\em left} $b_n$
module $b_n[d><d]$ (say) is thus
\[
M_d = \bigoplus_{d' \in R'_d} b_n [d'><d'] \;\; \in \;\; b_n^{h-2}
%.
\]
%(this is a slight abuse of notation as the right hand side
(we could make the sum over all $R_d$, but then it is not a direct sum
- the summands as written here may overlap at index $<(h-2)$).

 The bimodule subquotient $b^h_n/b^{h-2}_n$ of
$b_n$ decomposes via the diadic
structure as a direct sum of
%isomorphic
left modules, denoted
\begin{equation}
\label{eX}
{}_{b_n} \left( b^h_n/b^{h-2}_n \right) = \bigoplus_{t \in \{ e,f \} }
\left(
   \bigoplus_{d \in R^{n,h}_t}
\left( \left( b_n [d><d]  \right) \;
/ M_d    \right)    %\left( \bigoplus_{d'} [b^{k-2}><d'] \right)
\right)
\end{equation}
where all the summands inside the second sum are isomorphic.

The half diagrams
of given index $h$ and type $t$ ($t=e,f$) thus
provide a basis for representations of $b_n$, with action and
inner product both defined by the equation
\begin{equation}
\label{inner}
[a><b] \; [c><d] = <b|c> \; [a><d]
\hspace{1in}
(mod. \; b^{h-2}_n)
{}.
\end{equation}
In other words the `ket' (`bra') vectors give a basis for left
(right) modules (which
in our pictures means that the diagrams act from the top (bottom)).
Abusing symmetric group notation \cite{James} we call the corresponding left
modules Specht modules -
i.e. for $a \in R^{n,h}_t$
\[
S^{n,h}_t \sim b_n \; [a><a] \; / \; M_a
%.
\]
(independently of which $a \in   R^{n,h}_t$ is chosen).

\begin{de} For index $h$ %an integer
and type $t$  we
 define
a Gram matrix $g^{n,h}_t$ for the inner product by $(g^{n,h}_t)_{ab}=<a|b>$.
\end{de}
For example, with $n=3, h=1$ and a suitable order of the bases
(see table~\ref{pascal} below)
\[
\begin{array}{ccc}
g^{n,h}_f =
\left(
\begin{array}{ccc}
x&1&0
\\
1&y_f&0
\\
0&0&y_e
\end{array}
\right)
&
\hspace{.5in}
&
g^{n,h}_e =
\left(
\begin{array}{ccc}
y_f&0&0
\\
0&y_e&1
\\
0&1&x
\end{array}
\right)
\end{array}
{}.
\]
Noting that $g^{n,h}_t$ is symmetric, and non-singular for
$q,q'$ indeterminate, and that $[a><a]$ is
(up to normalisation) a primitive
idempotent in the apropriate subquotient, it follows that these
representations are inequivalent irreducible for $q,q'$ indeterminate.

There is a natural inclusion of
$b_n \subset b_{n+1}$ (add a propagating line
on the right). Hence
\begin{pr}[Induction/Restriction Diagram]
\label{pr2}
\label{P}
The dimensions, induction and restriction rules and bases for these
Specht modules are given by
%representations
%may be computed iteratively as follows.
%
%We then have
the following
generalised Pascal triangle for the iterative
construction of  `ket' (or
`bra') basis states. Starting with $n=1$ on the top layer:
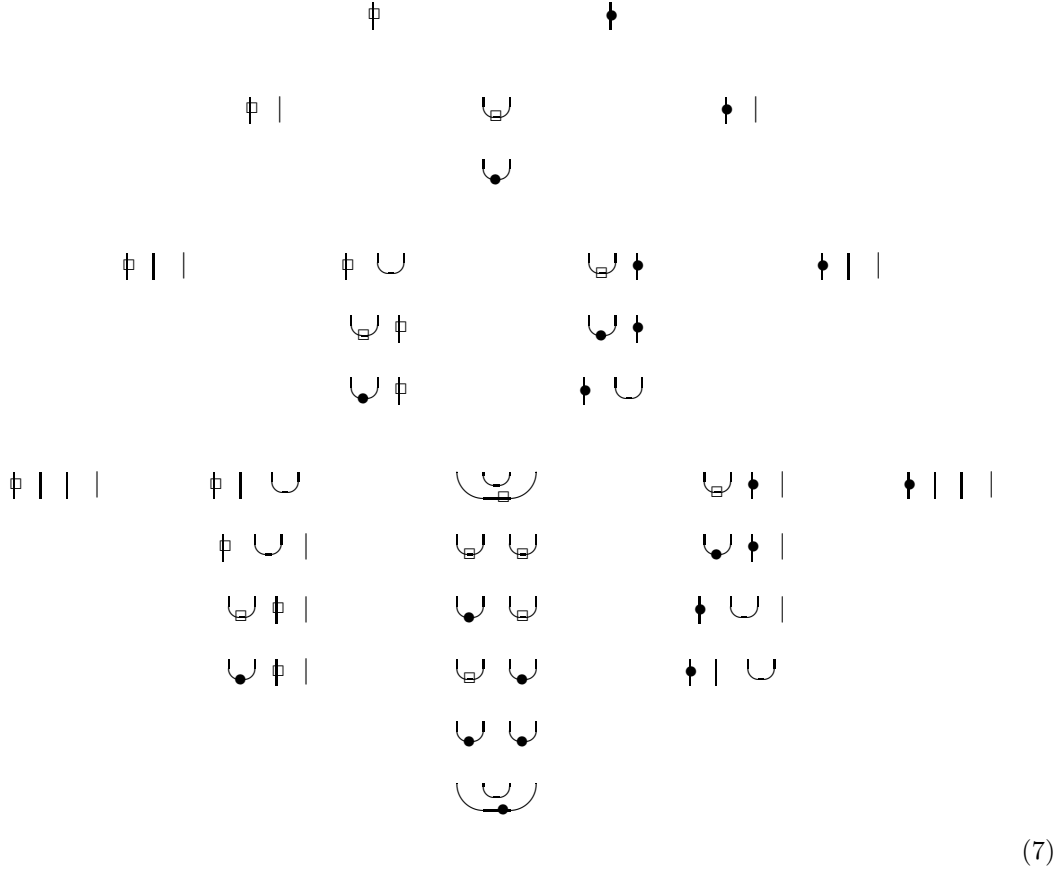
\[
\begin{array}{cccccccccccc}
&&&&\sq&&\bl
\\ \\
&&&\sq |&&\Cus&&\bl |
\\
&&&     &&\Cub
\\ \\
&&\sq \li | && \sq \Cup && \Cus \bl && \bl \li |
\\
&&        && \Cus \sq    && \Cub \bl
\\
&&        && \Cub \sq    && \bl \Cup
\\ \\
&\sq \li \li |&&\sq \li \Cup && \scus
                       && \Cus \bl | && \bl \li \li |
\\
&       &&\sq \Cup | && \Cus \Cus && \Cub \bl |
\\
&       &&\Cus \sq | && \Cub \Cus && \bl \Cup |
\\
&       &&\Cub \sq | && \Cus \Cub && \bl \li \Cup
\\
&       &&           && \Cub \Cub
\\
&        &&          && \scub
\\
\end{array}
\]
\begin{equation}
\label{pascal}
\end{equation}
and so on.
\end{pr}
Note that each new basis (with index $h$, say) is obtained by taking the
elements of the immediately
above left and right bases (which, unless $h=0$, are of
indices $h \pm 1$), adding a new line on the right of
each element (giving diagrams with indices $h$ and $h+2$), and then
in the index $h+2$ cases
connecting the new line to a previous line to
form a loop  (and so reduce the number of
propagating lines from $h+2$ to $h$).
Note that such a  connection is unique - as no propagating line may be
trapped by a cup, the rightmost propagating line must be used.

{\em Proof:} From the diagrams, or as follows....

%where the second sum is over diagrams $d' \in R^{n,k-2}$ obtained by taking
%any two (adjacent) propagating lines from $d$ and joining them to
%make a non-propagating line.

Let $[d_n><d_n]$ denote the inclusion of
$[d><d]$ in $b_{n+1}$ (i.e. $d_n$ has one extra propagating line
on the right, making $k+1$ altogether).
Then each of the summands in equation~\ref{eX} induces a
left $b_{n+1}$-module
\[
b_{n+1}[d_n><d_n] \left/
\bigoplus_{d' \in R'_d} b_{n+1}
 [d'_n><d'_n]    \right.
\]
In $b_{n+1} $
the quotient subspace sum is not quite over all of $R_{d_n}$,
since this includes the case in which the rightmost line
becomes looped back into the next such line (let's call it $d''$,
with $((h+1)-2)$ propagating lines).
Consequently, as a vector space the induced module from the index
$k$ Specht module is
a
direct sum of an index $k+1$ and an index $k-1$ Specht module -
\[
\left(
\left( \left( b_{n+1} [d_n><d_n] \right)
+ \left( b_{n+1} [d''><d''] \right) \right) \;
\left/
\bigoplus_{d' \in R'_d} b_{n+1}
 [d'_n><d'_n]    \right.
\right)
\hspace{2in}
\]
\[
\hspace{2in}
\cong
\left( \left( b_{n+1} [d_n><d_n] \right)
\left/
M_{d_n}  \right.
\right)
+
 \left( \left(
b_{n+1}  [d''><d''] \right)
\left/
M_{d''}   \right.
\right).
\]
%
%PAUL WHAT DO ... STAND FOR?

Restriction follows immediately from the diagrams.
QED.

\vspace{.1in}

This picture gives the generic structure of the algebra. To determine
the exceptional structures we note the following
\begin{pr}
For $x\ne 0$ and $n > 2$
\[
U_{n-1} b_n \; U_{n-1} \sim b_{n-2} U_{n-1}
\]
is an isomorphism of unital algebras.
\end{pr}
{\em Proof:}
Compare $U_{n-1} B^n_{\bullet}
U_{n-1} $ and
$ U_{n-1} B^{n-2}_{\bullet}$.
This has a standard corollary \cite{Green,Cline,Martin}
\begin{co}
\label{co3.1}
There exist functors on the categories of left modules
\[
\left( b_{n-2}-mod \right)
\stackrel{G}{\longrightarrow}
\left( b_{n}-mod \right)
\stackrel{F}{\longrightarrow}
\left( b_{n-2}-mod \right)
\]
such that $FG$ is the identity map, and
$GF(b_n)=b_n \; U_{n-1} \; b_n$.
\end{co}

The kernel of $GF$ determines the extent to which it fails to be an
isomorphism of categories.
 But this kernel is just
$b_n/b_nU_{n-1}b_n \sim e+f$ by
proposition~\ref{pr1}
(i.e. $b_n U_{n-1} b_n = b^{n-2}_n$), so exactly two {\em simple} modules are
missed in treating $b_{n-2}-mod$ as $b_n-mod$.
Now $b_1=\C \; e+ \C f$ and $b_2$ has three simple modules by
explicit computation.
It follows that the Pascal
diagram above gives bases for a complete list of
generic irreducible representations, since there two new nodes
appear in going from level $(n-2)$ to level $n$.

It also folows that at each level $n$
the only morphisms between
modules with a trivial image at
level $(n-2)$ are those
involving $b_n/b_{n-2} \sim
e \oplus f$. Therefore we can
build up the details of the
exceptional structure by looking
at these morphisms at each level, and adding them to the (known)
morphisms from level $(n-2)$.
The new morphisms can be determined from the zeros of the Gram matrices
$g^{n,k}_t$,
together with Frobenius reciprocity.

In what follows we %will sometimes
use the symmetry between the roles
of types $e$ and $f$. For generic type $t$ we will
then use $t'$ for the {\em other}
type.
It follows from the construction of bases in table~\ref{pascal}
that (where $||$ stands for determinant)
\begin{equation}
\label{ef}
|g^{2,0}_t|=y_f y_e
,
\end{equation}
$|g^{3,1}_t| =y_{t'}(y_t x -1)$,
%(here $t'$ means the other type!),
and for $n \ge 4$
\[
|g^{n,n-2}_e|
=
x \; |g^{n-1,n-3}_e| - |g^{n-2,n-4}_e|
\]
and similarly for the type $f$ cases. This is a recursion familiar
from the Temperley-Lieb case \cite{book}.
%If we treat $y_{t'}$ as an indeterminate and consider fixed $y_t$ (say)
%then $x$ is indeterminate and $|g^{n,n-2}_t|$ is linear in $y_t$.
%It has exactly one zero, at $y_t=P_{n-1}(x)/P_n(x)$ (c.f. \cite{book}),
%corresponding to the appearance of a one dimensional invariant subspace
%in $S^{n,n-2}_t$ isomorphic to $S^{n,n}_t$.

If we regard $|g^{n,n-2}_e|$ as a polynomial in $y_e$ then there are
various points at which it has (typically order 1) zeros. These
correspond to the occurence of a non-trnivial algebra homomorphism
at level $n$
\[
0 \rightarrow S^{n,n}_e \rightarrow S^{n,n-2}_e
{}.
\]
For example equation~\ref{ef} has a zero at $y_e=0$
corresponding to
\[
\bl | \;\;
\;\;\; \mapsto
\;\;\; \Cub
\]
(and similarly with the roles of $e$ and $f$ interchanged).
Frobenius reciprocity then determines a cascade of morphisms at
higher level, $m \; (>n)$ say, which exhaust the morphisms involving
$S^{m,m}_e$, the one
dimensional module based on $e$ (similarly $f$). All other
morphisms follow by corollary~\ref{co3.1}.

These morphisms determine the structure of the algebra. In
particular the dimensions of irreducibles may be determined by
a sequence of subtractions of dimensions of invariant subspaces.
This we now discuss in more details.

\subsection{The exceptional cases}

Introducing the standard sequence of polynomials
\begin{equation}
P^1_t=1,\ P^2_t=\frac{y_t}{x},\ P^{n}_t=P^{n-1}_t-x^{-2}P^{n-2}_t
\end{equation}
one finds therefore
\begin{equation}
|g_t^{n,n-2}|=\frac{y_{t'}}{x}x^n\ P^{n}_t
\end{equation}
It is useful to parametrize
\begin{equation}
y_e=\frac{q-q^{-1}e^{2i\eta}}{1-e^{2i\eta}},\ y_f=\frac{q^{-1}-q
e^{2i\eta}}{1-e^{2i\eta}}
\end{equation}
in which case the recursion relation is solved by
\begin{equation}
P^{n+1}_e=x^{-n}\frac{q^{-n}e^{2i\eta}-q^n}{e^{2i\eta}-1}
\end{equation}
and
\begin{equation}
P^{n+1}_f=x^{-n}\frac{q^{n}e^{2i\eta}-q^{-n}}{e^{2i\eta}-1}
\end{equation}
The zeroes of these polynomials occur at values
\begin{equation}
\label{eq6}
\eta=\pm n\gamma   %\hbox{ mod }
        + m \pi  \hspace{1in} \mbox{$m$ integer}
\end{equation}
 respectively, where we set
$q=e^{i\gamma}$.

For given $\eta,\gamma$ there are various possibilities to consider:

Firstly if %$\eta/\gamma \not \in Z$
there are no solutions to equation~\ref{eq6} then the algebra is
semi-simple (generic) for all $n$. This is because
by corollary 3.1 the first  occurence
(as we increase $n$) of an invariant subspace in a generic
indecomposable module  $S^{n,h}_t$ (for some $h$)
must give a homomorphism from the $t$-trivial
module
(that is, if $t=e$ say,
the {\em rightmost} module in level $n$ of the Bratteli
diagram)
\begin{equation}
0 \rightarrow S^{n,n}_t \rightarrow S^{n,h}_t
{}.
\end{equation}
But since $S^{n,h}_t$ restricts to $S^{n,h-1}_t + S^{n,h+1}_t$
Frobenius reciprocity \cite{mathfool}
implies a morphism from one of these to
$S^{n-1,n-1}_t$ - a contradiction unless $h=n-2$.

\vspace{.1in}

Secondly if equation~\ref{eq6} is satisfied for some sign and a
pair of integers $m,n=m_c,n_c$,
but there are no other solutions (i.e. $q$ not a root of unity),
then by the same argument as above this signals the {\em first}
occurence of an invariant subspace in a generically irreducible
module. All subsequent homomorphisms are determined by Frobenius
reciprocity and corollary 3.1. In particular all modules $S^{n,h}_t$
with $h>n_c$ remain irreducible (suppose there is a first one which
does not, then it must have an invariant subspace isomorphic to
$S^{n,n}_t$ by co.3.1, but then by Frobenius reciprocity there must
either be an earlier one with an
invariant subspace or else $h=n-2$ - either way we have a contradiction).
Further, for each positive $l$ such that $n_c+l \leq n$ we have
\begin{equation}
\label{meq2}
0 \rightarrow S^{n,n_c+l}_t \rightarrow S^{n,n_c-l}_t
\end{equation}
where the cases $n_c+l <n$ are given
by corollary 3.1, and for $l=n-n_c$ the morphism follows by Frobenius
reciprocity, and there can be no other morphisms. Note that when
$n_c-l <0$ we have $S^{n,n_c-l}_t = S^{n,l-n_c}_{t'}$
(a notational convenience from the generic Bratteli diagram).

In this case the dimensions of the new irreducibles may be computed by
subtracting $dim(S^{n,n_c+l}_t)$ from the generic
dimension, or noting
that {\em for these irreducibles}, call them $I^{n,h}_t$, the induction
and restriction rules are the same as before, except that
$I^{n,n_c-1}_t$ restricts (where defined) to  $I^{n-1,n_c-2}_t$
(and not $I^{n-1,n_c-2}_t+I^{n-1,n_c}_t$ as usual); and
$I^{n,n_c}_t$ restricts (where defined) to
$I^{n-1,n_c-1}_t +
I^{n-1,n_c+1}_t +
I^{n-1,n_c+1}_t$.

\vspace{.1in}

Finally, if there is a solution to equation~\ref{eq6}
and $q$ is a root of unity then
$\gamma / \pi=m_1/m_2$ for some coprime integers $m_1,m_2$ and
there is another solution for each integer $m$ such that
 $(m-m_c) \pi / \gamma \in Z$ (i.e. $m=m_c$ mod $m_1$),
at $n= | n_c + k m_2 |$     %(positive
(integer $l$).
%This is just when $n=n_c$ mod $m_2$.
Without loss of generality let us assume that the lowest $n$ solution,
at $n=n_c$,
occurs when the positive sign occurs in equation~\ref{eq6}, i.e. a
solution to $P^{n+1}_e=0$. The next
(or equal) lowest $n$ solution will be
to
 $P^{n+1}_f=0$, at $n=m_2-n_c$, and so on.

As before, for $n \le n_c$ the algebra is generic. At $n=n_c+1$ the only
morphism is as in equation~\ref{meq2}
(with $l=1$). n
For $n \le  n_c + m_2$ the structure is as
in the single solution case above, except that at $n=m_2-n_c+1$
morphisms involving the $f$-trivial module (and their descendents
by corollary 3.1)
\begin{equation}
\label{meq3}
0 \rightarrow S^{n,m_2-n_c+l}_{t'} \rightarrow S^{n,m_2-n_c-l}_{t'}
\end{equation}
begin to appear, so that some modules will have two invariant
subspaces (one coming from the $e$ and one from the $f$ side).

At $n=n_c +m_2$ there is another solution on the $e$ side. This
requires a refinement to the Frobenius reciprocity argument to
derive the new structure. The morphism from $S^{n_c+m_2,n_c+m_2}_t$ to
 $S^{n_c+m_2,0}_t$ is still forced, but due to the
new solution there is also one from
$S^{n_c+m_2,n_c+m_2}_t$ to
$S^{n_c+m_2,n_c+m_2-2}_t$. Frobenius reciprocity then forces a new
series of morphisms, together with corollary 3.1 altogether giving
\begin{equation}
\label{meq4}
0 \rightarrow S^{n,m_2+n_c+l}_{t} \rightarrow S^{n,m_2+n_c-l}_{t}
\end{equation}
(all appropriate positive integer $l$)
by an anlogous argument to that above.
It is convenient to think of these morphisms as
corresponding to `reflections' in
the vertical `critical line' in the Bratteli diagram begining at
the $S^{n_c+m_2,n_c+m_2}_t$ position.
In these terms the earlier sets of morphisms
are then given by `reflections' in the vertical lines begining at
$S^{m_2-n_c,m_2-n_c}_{t'} $ and $S^{n_c,n_c}_t$ respectively.
Every time a new solution appears
(at $|n_c + k m_2|$, on the $t$ side for $k$ positive integer, and
on the $t'$ side for $k$ negative integer ($t=e$ here in case of positive
sign in equation~\ref{eq6}))
a new set of such morphisms is initiated. For
example $n_c +k m_2$ gives
\begin{equation}
\label{meq5}
0 \rightarrow S^{n,k m_2+n_c+l}_{t} \rightarrow S^{n,k m_2+n_c-l}_{t}
\end{equation}
(all appropriate positive integer $l$).
Each time a new set of morphisms is initiated in this way at some level
$n$, the modules involved in the morphisms of previously initiated
(lower $|k|$) sets occuring at that level (and subsequent levels)
become involved in {\em chains} of morphisms (that is,
the domains are also ranges of morphisms from the {\em new}
set). In this general case the appropriate morphism is
\begin{equation}
\label{meq6}
0 \rightarrow I^{n,k m_2+n_c+l}_{t} \rightarrow S^{n,k m_2+n_c-l}_{t}
\end{equation}
where $I^{n,k m_2+n_c+l}_{t}$ is simple, i.e.
\[
I^{n,k m_2+n_c+l}_{t}=S^{n,k m_2+n_c+l}_{t}/I_{\bullet}
\]
($I_{\bullet}$ the maximal proper invariant subspace of
$S^{n,k m_2+n_c+l}_{t}$).
Indeed, since $I=S$ {\em far enough out}
in the Bratteli diagram we might as well write all our
morphisms as in equation~\ref{meq6}.

In general there are many morphisms in and out of each generically
irreducible module,
but all $e$ type morphisms map from right to left, and
all $f$ type from left to right (corollary 3.1). Thus by working in from the
edges in a suitable order all
the dimensions of simple modules can be computed by subtractions of
{\em known}  dimensions of
invariant subspacesfor the  generic case.
The induction and restriction rules for the new simple modules
may be worked out directly from this.

Alternatively the dimensions
of simple modules may be thought of in terms of subsets
of walks from the top of the diagram to the position of
the module in
question. The simple dimension for a module between two
critical lines
{\em on the same side} is the number of walks
which do not
touch the innermost line on the {\em other} side, and
which do not
touch the outer line of the two after the last time they they touch the
inner one. For a module {\em on} a critical line all walks which
never touch the first critical line on the other side are allowed.
For a module between the two innermost lines
(one on each side) only walks which
never touch either line are allowed.
Thus for this innermost sector in particular
we may summarize:

If there is an integer $n$ (the smallest) for which $P^{n+1}_e$
vanishes we may truncate the Bratteli diagram on the right (e part) to
keep only connectivities $h\leq n-1$.
Similarly if
 $P^{n'+1}_f$ vanishes we truncate it  on the left to keep only connectivities
$h\leq |1-n'|$. If there is a
pair of integers $n,n'$ such that $P^n_e$ and $P^{n'}_f$ vanish, this implies
that $q$ is a $(n+n')^{\mbox{th}}$ root of unity. In this case the Bratteli
diagram may be truncated on both sides.

Our results in this
sector are in correspondence with the ones of Levy \cite{Lev}
who studied an  algebra $Y(\tau,a,b,c,N)$ generated by
$1,e_1,\ldots,e_{N-1};x_1$ with usual TL relations $e_i^2=e_i,\
e_ie_{i\pm 1}e_i=\tau e_i,\ [e_i,e_j]=0,\ |i-j|\geq 2$ and
$x_1^2=bx_1+c,e_1x_1e_1=ae_1$. Besides a simple rescaling and renaming
 of the various generators,
the correspondence with us is $\tau=x^2$ and
\[
e=\frac{\mu x_1+1}{\mu b+2}, y=x\frac{\mu a+1}{\mu b+2}
\]
where
\[
\mu=\frac{b\pm\sqrt{b^2+4c}}{2c}
\]

Our results agree with those of \cite{Lev}
in this sector, but we believe the representation
theory is much more transparent our way, this belief being reinforced by
the fact that we obtain the {\em whole} structure, not just the
innermost part.

\subsection{Discussion}
Note that $e$ can be braid translated
(conjugated by the usual braid generator $g_1=1-qU_{1}$
of the Temperley-Lieb algebra) so that the blob appears in other places
besides the first strand. Leaving it on the first strand is just a
prescription to ensure linear independence of diagrams.
The blob can be thought of as a trick for introducing
%an extra
%degree of freedom required for building
a cohomological `seam' into the
system - the first step in generalizing to periodic boundary
conditions. As such only one blob is required, but the seam can
occur anywhere in the chain.

\section{Application to the periodic Temperley Lieb algebra}

We now wish to apply the above results to the study of the periodic Temperley
Lieb algebra $T_{\hat{A}_{n-1}}$\cite{PS90,Lev1}\footnote{Recall that the
$\hat{A}_n$ diagram has $n+1$ vertices}. Rename first the generators
of $T_{2n-1}(q)$ as $U_{1.}, U_{12}, \ldots, U_{n.}$. Then $T_{\hat{A}_{n-1}}$
(denoted simply by $T$ in the following)
is the unital algebra over the complex numbers
 generated by these generators and
an additional one -
 $U_{n1}$ - that satisfies the relations
\[
U_{n1}^2=xU_{n1}
\]
\[
 U_{n.}U_{n1}U_{n.}=U_{n.}, \ U_{1.}U_{n1}U_{1.}=U_{1.},
\ U_{n1}U_{n.}U_{n1}=U_{n1}, \ U_{n1}U_{1.}U_{n1}=U_{n1}
\]
\begin{equation}
[U_{n1}, U_{i.}]=0\  i\neq 1\hbox{ or }n,
[U_{n1},U_{i,i+1}]=0\label{eq:period}
\end{equation}
It is an infinite dimensional algebra.

Setting $I_0=\prod_{i=1}^{n}(U_{i.}/\sqrt{Q})$
consider first the left ideal $TI_0$. Recall from \cite{MS} that all
irreducible
representations may be found by considering the quotients
\begin{equation}
\left(U_{1.}U_{2.}\ldots U_{n.}\right)U_{12}U_{23}\ldots U_{n1}
\left(U_{1.}U_{2.}\ldots U_{n.}\right)=\alpha
\left(U_{1.}U_{2.}\ldots U_{n.}\right)\label{eq:quotient}
\end{equation}
for some parameter $\alpha$. $TI_0$ modulo (\ref{eq:quotient}) is
indecomposable. It has a natural basis of words in the generators, and we call
the representation induced from this basis ${\cal T}_n{z\choose 0}$ where we
have set

\begin{equation}
\alpha=(z^{1/2}+z^{-1/2})^2
\end{equation}

There is an algebra homomorphism from
the periodic Temperley-Lieb algebra        %(i.e. with additional generator
%%$U_0=U_n$)
into $T_{2n-1}(q)$ given by
\[
P: T_{\hat{A}_{n-1}}(q) \rightarrow T_{2n-1}(q)
\]
\[
P: U_{i.} \mapsto U_{i.}   \hspace{1in} (i=1,2,...,n-1)
\]
(similarly $U_{i \; i+1}$) and,
recalling  the "braid translator" introduced in \cite{MS}, by
\begin{equation}
P:U_{n1} \mapsto \left(\prod_{i=1}^{n-1}g_{i.}g_{ii+1}\right)^{-1}
U_{1.}\left(\prod_{i=1}^{n-1}g_{i.}g_{ii+1}\right)
\end{equation}
where $g^{\pm 1}=1-q^{\pm 1}U$. It is easy to check that
this realization of $U_{n1}$ satisfies the
relations  (\ref{eq:period}). Moreover  (\ref{eq:quotient}) holds with
$\alpha=x^2$. Hence we were able in \cite{MS} to induce representations of $T$
with that particular value of $\alpha$ from representations of $T_{2n-1}$.

We can now  generalize the braid translator by considering blob
and squares decorations.
This should allow an algebra homomorphism from the periodic
alegbra into the blob algebra.
For given parameters $x$ and $y_f$ consider
now trying to build a homomorphism of the form
\begin{equation}
U_{n1} \mapsto (a f+1)\left(\prod_{i=1}^{n-1}g_{i.}g_{ii+1}\right)^{-1}
U_{1.}\left(\prod_{i=1}^{n-1}g_{i.}g_{ii+1}\right)(b f+1)
\end{equation}
It is easy to check that for any $y_f$ this
%realization of the
generator satisfies the relations
(\ref{eq:period}) provided the following conditions hold
\begin{eqnarray}
a+b+ab&=&0\nonumber\\
aq^{-1}+bq-y_f(a+b)&=&0
\end{eqnarray}
where the first equation follows from idempotency of $U_{n1}/x$ and the second
from the relations invloving three $U's$. Beside the trivial solution $a=b=0$
used in \cite{MS} another possibility is
\begin{eqnarray}
a&=&\frac{q-q^{-1}}{q^{-1}-y_f}\nonumber\\
b&=&\frac{q^{-1}-q}{q-y_f}
\end{eqnarray}
In that case (\ref{eq:quotient}) holds with
\begin{equation}
\alpha=x^2+\frac{x^2-4}{y_f^2-xy_f+1}y_f(x-y_f)
\end{equation}
Setting
\begin{equation}
y_f=\frac{q-q^{-1}e^{2i\eta}}{1-e^{2i\eta}}
\end{equation}
one finds
\begin{equation}
z=\hbox{exp}(2i\eta)
\end{equation}
%234z          ?? NAN DESU KA, SALEUR-SAN? I ASSUME IT IS A GLITCH!
%
(all choices of phases for $z,q$ give isomorphic results up to $e,f$
interchange).
Therefore we can establish an isomorphism between ${\cal T}_n{z\choose 0}$
as introduced in \cite{MS} (sec. 4.3) and  representations of
the blob algebra $S_f^{2n,0}(y_f)$.
{}From the results of the previous paragraph we deduce immediately  that
 ${\cal T}_n{z\choose 0}$ is irreducible for $z$ generic,
with dimension the number of
paths of $2n$ steps from the origin to the point of horizontal coordinate zero
on the Pascal triangle (here for convenience we use
horizontal coordinates that are equal to the number of connectivities on the
$e$ side, and minus it on the $f$ side)
, ie $C_{2n}^n$. It is reducible when
$\hbox{exp}i\eta=q^{n_c}$ for $n_c$ an integer, that is $z=q^{2k}$. By
symmetry we can restrict to the case $n_c$ positive. The representation
 contains then
an irreducible component with dimension the number of paths
with same characteristics but on a Pascal triangle that is truncated on the
left to include only points with horizontal coordinates greater or equal to
$1-n_c$, ie $C_{2n}^n-C_{2n}^{n-n_c}$.
 In \cite{MS} since we restricted to the case without blob, we could use
the braid translator only in the case $n_c=1$. This recovers the results
of prop.16 in
\cite{MS}.

The process generalizes to the case $TI_h/TI_{h-1}T$ where
$I_h=\prod_{i=1}^{n-h}(U_{i.}/\sqrt{Q})$ (that is the sector with $2h$
propagating lines. Notice that connectivities as they
have been defined so far are {\em half} of the connectivities defined in
\cite{MS} where they referred to ``clusters'' rather than ``boundaries'').
The relevant quotient relations are
obtained by taking the word $I_h$,  rotating the top  once around the cylinder
clockwise holding the bottom fixed, and equating this new word with
$\alpha_hI_h$. The same result with $\alpha_h^{-1}$ holds then for
counterclockwise rotation. In the case $h=1$ an additional quotient has to be
taken
\begin{equation}
\left(U_{1.}U_{2.}\ldots U_{(n-1).}\right)U_{12}U_{23}\ldots U_{n1}
\left(U_{1.}U_{2.}\ldots U_{(n-1).}\right)=0
\end{equation}
Quotienting $TI_h/TI_{h-1}T$
by these relations
one obtains  the representation ${\cal T}_n{\alpha_h\choose h}$.
On the other hand from braid translating the blob algebra one gets the same
relations with the parameter
\begin{equation}
\alpha_h=q^{2h}\hbox{exp}(2i\eta)
\end{equation}
so
we have isomorphism with $S_f^{2n,2h}(y_f)$.
The results of \cite{MS} immediately follow. For $\alpha_h$ generic,
${\cal T}_n{\alpha_h\choose h}$ is
 irreducible with dimension the number of
paths of $2n$ steps going from the origin to the point of horizontal coordinate
$2h$ on the Pascal triangle, ie $C_{2n}^{n-h}$. The representation is
reducible for $\alpha_h=q^{2k}$ where $k=h+n_c,\ n_c=1,\ldots, n$.
In that case it
contains an irreducible component $\rho_n{h+n_c\choose h}$
of dimension the number of paths with same
characteristics but on a diagram truncated on the left to contain only points
with horizontal coordinates greater or equal to $1-n_c$, ie
$C_{2n}^{n-h}-C_{2n}^{n-h-n_c}$ (Notice this
  coincides with the number of paths with same characteristics but on a
diagram truncated on the right to contain only points of horizontal coordinate
lower or equal to $2h+n_c-1$, as well
as the number of paths
of $2n$ steps on a half Pascal triangle (with only positive coordinates)
that go from a point of horizontal
coordinate $n_c-1$ to a point of horizontal coordinate $2h+n_c-1$ see figure
 1).
This recovers the results of prop.18 in \cite{MS}.

The connection  with the blob algebra allows us as well to study the
representation theory of $T$ when $q$ is a root of unity. This was not
straightforward using the Gram determinants results of \cite{MS} due to the
existence of multiple zeroes. Suppose $m_2$ is the
smallest integer such that $q^{m_2}=\pm 1$
. Then $\rho_n{h+n_c\choose h}$
is further
reducible. It contains an irreducible component $\rho_{ab}(n)$
 where $a=n_c-1, b=2h+n_c-1$ ($0\leq a,b\leq m_2-2$)
of dimension the number of paths of $2n$
steps that go from the origin to the point of horizontal coordinate $2h$ on a
Pascal triangle that is truncated on the left and on the right so as to include
only points of horizontal coordinate greater or equal to $1-n_c$ and smaller or
equal to $p-2+1-n_c$ (or horizontal coordinate
greater or equal to $2-p+2h+n_c-1$
and smaller or equal to $2h+n_c-1$).
This is as well the number of paths of $2n$ steps on a half
Pascal triangle truncated on the right to contain points of coordinate smaller
or equal to $p-2$, that
 go from a point of horizontal
coordinate $n_c-1$ to a point of horizontal coordinate $2h+n_c-1$ (see
figure 2). The explicit
expression of this dimension is, for $(a,b)\neq (m_2/2-1,m_2/2-1)$
\begin{equation}
\hbox{dim}\rho_{ab}=\sum_{i\in Z}{n-\frac{a-b}{2}+im_2\choose 2n}-
{n-\frac{a+b}{2}-1+im_2\choose 2n}
\end{equation}n
where the sum truncates for negative arguments in the binomial coefficients.
When  $(a,b)= (m_2/2-1,m_2/2-1)$  the dimension is half of the above
expression. The representations  $\rho_{ab}$ and $\rho_{m_2-a-2,m_2-b-2}$ are
isomorphic by $e,f$ interchange symmetry.

These results were conjectured first in \cite{PS90}.
In this latter reference, the decomposition of the reducible
representations ${\cal R}$ of $T$
provided by solid on solid models on Dynkin diagrams ${\cal D}$ \cite{P87}
were also given.
Recall for instance in the simplest case of the Ising model ($A_3$)
and the 3 state
 Potts model ($D_4$)
\begin{equation}
{\cal R}^{A_3}=\rho_{00}+\rho_{11},\ {\cal
R}^{D_4}=\rho_{00}+2\rho_{22}+\rho_{04}
\end{equation}

\section{Conclusion}

As well as giving a complete analysis of the blob algebra
this paper  completes the study of the representation theory of
the periodic Temperley Lieb algebra \cite{PS90,Lev1,MS}. The analogy
between representations of this algebra and those of left - right Virasoro
algebra goes actually further than the ``minimal set'' that was mainly
discussed so far \cite{PS90}:
representations that
are not in the innermost part of the Bratteli diagram can as well be
put in correspondence
with representations of the Virasoro algebra
that are ``outside'' the minimal Kac table. This is easily checked as
in \cite{PS90} for instance by calculating traces of the physical
hamiltonian in the continuum limit and comparing them with the known
Virasoro characters.

An interesting physical question is whether there are lattice models of
restricted solid on solid type that use only the outside representations, the
way the Andrews Baxter Forrester \cite{ABF84}
models use only the innermost part
of the Bratteli diagram?. By conformal invariance analogy one
expects there are no such models that lead to modular invariant
partition functions. This is because, for a conformal field theory,
if one representation outside the minimal Kac table appears, then
all the ones inside must appear as well by the effect of modular
transformations \cite{DFSZ}. Indeed, it is easy to build SOS models
that use only outside representations by a limiting process
on the interacting round a face form of the Temperley Lieb generators. But
the constraint that the paths must not touch the outer line after the last time
they touched the inner one makes such model very different in space and
time direction, and likely enough cannot lead to a modular invariant
partition function.

 An interesting question concerns  the nature and properties
of   ``physical'' representations
of the blob
 algebra. For instance the vertex model representation of the Temperley
lieb algebra is well known \cite{B}, with basis provided by $n$ tensored
copies of $\C
^{\ 2}$ and $U$ matrices acting between two neighboring copies
($U_i$ acts on the $i^{th}$ and $(i+1)^{th}$ copies) as
the matrix
\begin{equation}
U=\left(\begin{array}{cccc}
0&0&0&0\\
0&q^{-1}&-1&0\\
0&-1&q&0\\
0&0&0&0
\end{array}\right)
\end{equation}
 It is
immediate then to find a representation of $e$ acting on the left most copy
by the matrix
\begin{equation}
e=\frac{1}{a+a^{-1}}\left(\begin{array}{cc}
a^{-1}&-1\\
-1&a
\end{array}\right)
\end{equation}
with the value
\begin{equation}
y_e=\frac{aq+a^{-1}q^{-1}}{a+a^{-1}}
\end{equation}
Note that this matrix breaks the `charge conservation'
property which allows the vertex model representation to be
immediately broken up into blocks ($q$-analogues of permutation blocks
in the symmetric group).
The use of this representation is not completely clear to us.

\begin{figure}
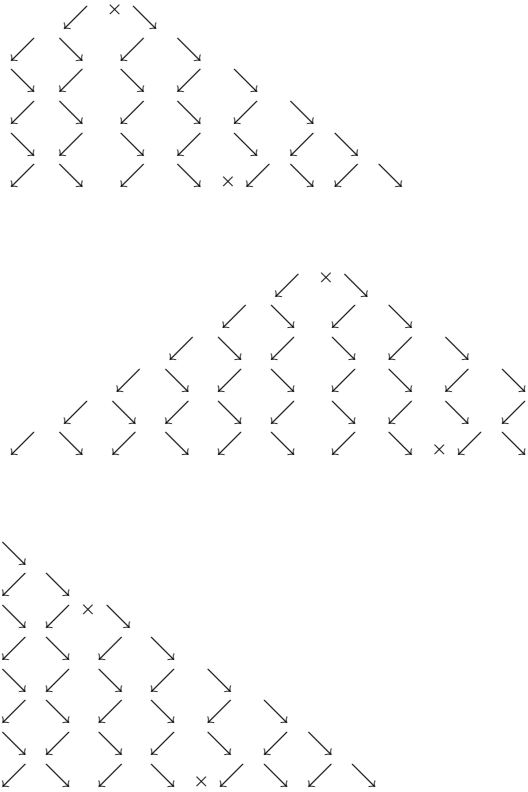


$\begin{array}{cccccccccccccccccccccccccccccc}
&\swarrow&\!\!^\times\searrow\\
\swarrow&\!\!\searrow&\!\!\swarrow&\!\!\searrow\\
\searrow&\!\!\swarrow&\!\!\searrow&\!\!\swarrow&\!\!\searrow\\
\swarrow&\!\!\searrow&\!\!\swarrow&\!\!\searrow&\!\!\swarrow&
\!\!\searrow\\
\searrow&\!\!\swarrow&\!\!\searrow&\!\!\swarrow&\!\!\searrow
&\!\!\swarrow&\!\!\searrow\\
\swarrow&\!\!\searrow&\!\!\swarrow &\!\!\searrow&\!\!_\times\swarrow&
\!\!\searrow&
\!\!\swarrow&\!\!\searrow
\end{array}$

\vspace{1cm}

$\begin{array}{ccccccccccccccccccccccccccccccccc}
&&&&&\swarrow&\!\!^\times\searrow\\
&&&&\swarrow&\!\!\searrow&\!\!\swarrow&\!\!\searrow\\
&&&\swarrow&\!\!\searrow&\!\!\swarrow&\!\!\searrow&\!\!\swarrow&\!\!\searrow\\
&&\swarrow&\!\!\searrow&\!\!
\swarrow&\!\!\searrow&\!\!\swarrow&\!\!\searrow&\!\!\swarrow&
\!\!\searrow\\
&\swarrow&\!\!\searrow&\!\!\swarrow&\!\!
\searrow&\!\!\swarrow&\!\!\searrow&\!\!\swarrow&\!\!\searrow
&\!\!\swarrow\\
\swarrow&\!\!\searrow&\!\!\swarrow &\!\!\searrow&\!\!\swarrow&
\!\!\searrow&\!\!\swarrow&\!\!\searrow&\!\!_\times\swarrow &\!\!\searrow
\end{array}$

\vspace{1cm}

$\begin{array}{ccccccccccccccccccccccccccccccccccccccc}
\!\!\searrow&\\
\!\!\swarrow&\!\!\searrow&\\
\!\!\searrow&\!\!\swarrow&\!\!\!\!^\times\searrow&\\
\!\!\swarrow&\!\!\searrow&\!\!\swarrow&
\!\!\searrow&\\
\!\!\searrow&\!\!\swarrow&\!\!\searrow
&\!\!\swarrow&\!\!\searrow&\\
\!\!\swarrow&\!\!\searrow&\!\!\swarrow&
\!\!\searrow&
\!\!\swarrow&\!\!\searrow\\
\!\!\searrow&\!\!\swarrow&\!\!\searrow&
\!\!\swarrow&
\!\!\searrow&\!\!\swarrow&\!\!\searrow\\
\!\!\swarrow&\!\!\searrow&\!\!\swarrow&
\!\!\searrow&
\!\!_\times\swarrow&\!\!\searrow&
\!\!\swarrow&\!\!\searrow\\
\end{array}$

\vspace{1cm}

\caption{
Three ways of counting the number of states in $\rho_{n=3}{k=4\choose
h=1}$. This number coincides with the number of paths of 6 steps on any
of the diagrams that go form the top (with coordinate 0) to the bottom cross (
with coordinate 2).}

\end{figure}

\begin{figure}
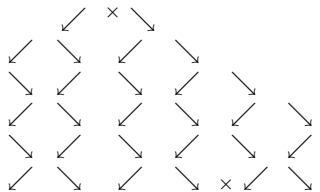


$\begin{array}{cccccccccccccccccccccccccccccccccc}
&\swarrow&\!\!^\times\searrow\\
\swarrow&\!\!\searrow&\!\!\swarrow&\!\!\searrow\\
\searrow&\!\!\swarrow&\!\!\searrow&\!\!\swarrow&\!\!\searrow\\
\swarrow&\!\!\searrow&\!\!\swarrow&\!\!\searrow&\!\!\swarrow&
\!\!\searrow\\
\searrow&\!\!\swarrow&\!\!\searrow&\!\!\swarrow&\!\!\searrow
&\!\!\swarrow\\
\swarrow&\!\!\searrow&\!\!\swarrow &\!\!\searrow&\!\!_\times\swarrow&
\!\!\searrow\\
\end{array}$

\vspace{1cm}

\caption{Same as the first diagram above, but  for $q$ a root of unity
with $m_2-2=6$}

\end{figure}

\vspace{.1in}

\noindent {\bf Acknowledgements} $\;\;$
One of us (PPM) would like to thank B W Westbury and D Levy for useful
discussions,
and to thank SERC and the Nuffield Foundation for financial support in
completing part of this project.
HS was supported by the Packard foundation.


\begin{thebibliography}{99}
\bibitem{McCoy} G. Albertini, S. Dasmahapatra and B.M. McCoy, Int.J.Mod. Phys.
A7 Suppl.1A (1992) 1;
F.Alcaraz, U.Grimm,
V.Rittenberg, Nucl. Phys. {\bf B316}, (1989) 735
\bibitem{B}R.J.Baxter, ``Exactly solved models in
 statistical mechanics'', (Academic Press, New York, 1982).
\bibitem{PS90}V.Pasquier, H.Saleur, {\sl Nucl. Phys.}, {\bf  330}, (1990) 523
\bibitem{Lev} D.Levy,
``The structure of the affine Hecke quotient underlying the
translation invariant $XXZ$ chain'', Tel-Aviv preprint TAUP 1986-92.
\bibitem{Lev1}D.Levy, Phys.Rev.Lett. 67 (1991) 1971.
\bibitem{MS} P.P.Martin and H.Saleur, "On an algebraic approach to higher
dimensional
statistical mechanics", preprint YCTP-P33-92.
\bibitem{BPZ} A.Belavin, A.Polyakov, Al.B.Zamolodchikov, Nucl. Phys. {\bf
B241}, (1984) 333
\bibitem{TL}H.N.V.Temperley and E.Lieb, Proc.R.Soc. A322(1971) 25.
\bibitem{Pen} R.Penrose, in ``Quantum theory and beyond'',
ed. T.Bastin, (Cambridge U., New York, 1971) p.151.
\bibitem{Westbury} B.W.Westbury, Manchester University 1990 Ph.D. Thesis
and unpublished 1991 notes
``On the Temperley-Lieb algebras''.
\bibitem{James} G.D. James and A. Kerber, ``The representation theory of the
symmetric group'',
(Addison-Wesley,
Reading, 1981).
\bibitem{Green} J.A.Green, ``Polynomial representations of $Gl_n$'',
Springer Lecture notes in Mathematics
 830 (Berlin, 1980).
\bibitem{Cline} E. Cline, B. Parshal and L. Scott, J. reine angew. Math. 391
(1988) 85.
\bibitem{Martin} P.P.Martin , ``Non-Planar Statistical Mechanics -
The Partition Algebra
construction'', Yale preprint YCTP-P34-92.
\bibitem{book} P.P.Martin,
``Potts Models and related problems in statistical mechanics'',
World Scientific, Singapore, 1991.
\bibitem{mathfool} see for example
P.M.Cohn, {\em Algebra} vol.2, (John Wiley, London, 1989).
\bibitem{P87} V.Pasquier, {\sl Nucl. Phys.}, {\bf B285}, (1987)162
; {\sl J.Phys. }, {\bf A20}, (1987) L1229
\bibitem{ABF84}G.E.Andrews, R.J.Baxter, P.J.Forrester, {\sl J.Stat.Phys.}, {\bf
35}, (1984) 193
\bibitem{DFSZ}P.Di Francesco, H.Saleur, J.B.Zuber, {\sl Nucl. Phys.}, {\bf
285}, (1987) 454
\end{thebibliography}
\end{document}